# Feature set optimization by clustering, univariate association, Deep & Machine learning omics Wide Association Study (DMWAS) for Biomarkers discovery as tested on GTEx pilot dataset for death due to heart-attack

Abhishek Narain Singh

Web: ABioTek www.tinyurl.com/abinarain

abhishek.narain@iitdalumni.com

## Abstract

Univariate and multivariate methods for association of the genomic variations with the end-or-endo phenotype have been widely used for genome wide association studies. In addition to encoding the SNPs, we advocate usage of clustering as a novel method to encode the structural variations, SVs, in genomes, such as the deletions and insertions polymorphism (DIPs), Copy Number Variations (CNVs), translocation, inversion, etc., that can be used as an independent feature variable value for downstream computation by artificial intelligence methods to predict the endo-or-end phenotype. We introduce a clustering based encoding scheme for structural variations and omics based analysis. We conducted a complete all genomic variants association with the phenotype using deep learning and other machine learning techniques, though other methods such as genetic algorithm can also be applied. Applying this encoding of SVs and one-hot encoding of SNPs on GTEx V7 pilot DNA variation dataset, we were able to get high accuracy using various methods of DMWAS, and particularly found logistic regression to work the best for death due to heart-attack (MHHRTATT) phenotype. The genomic variants acting as feature sets were then arranged in descending order of power of impact on the disease or trait phenotype, which we call optimization and that also uses top univariate association into account. Variant Id P1_M_061510_3_402_P at chromosome 3 & position 192063195 was found to be most highly associated to MHHRTATT. We present here the top ten optimized genomic variant feature set for the MHHRTATT phenotypic cause of death.

## Keywords

SNP – Single Nucleotide Polymorphism, SVs – Structural Variations, MLP – Multi-Layer Perceptron, DNN – Deep neural network,

DNA – Deoxyribonucleic acid, DIP – Deletion and Insertion Polymorphism, InDel – Insertion Deletion, DMWAS – Deep and Machine learning omics Wide Association Study, GWAS – Genome Wide Association Study, NGS – Next Generation Sequencing, ExhaustiveDNN – Exhaustive Deep Neural Network, TP – True Positive, TN – True Negative, FN – False Negative, FP – False Positive , CNV – Copy Number Variation, MLCSB – Machine Learning in Computational and Systems Biology, AUC – Area under curve, ROC – Receiver Operating Characteristic, PR-curve – Precision-recall curve, MHHRTATT - people who died of 'heart attack, acute myocardial infarction, acute coronary syndrome'

### Background

Genomes of individuals are said to be more than 99% similar. This small variation of less than 1% in the DNA accounts for vast amount of differences in endo-and-end phenotype and behavior of the person. Variations of single letters in nature, such as the letters A, T, G, C, N can be easily encoded, while numerical representation of variations of DNA of more than 1 letter need more complicated and logical methods. GWAS[12] has been used for univariate methods of association of these variations to end phenotype until now. A univariate method for association of the genomic variations with the end-or-endo-phenotype has been widely used through software tools such as snptest [23] and p-link [22]. Methods of multivariate GWAS where there are multiple phenotypes to associate with as dependent variables, which are claimed to perform better, have been suggested [24]. However, these associations still take one independent variable at a time for genome wide association, therefore are less stringent resulting in spurious results. We see lately that overall contribution of these loci to heritability of complex diseases is often less than 10% [13]. As pointed out from McClellan and King, Cell 2010[14]:

"To date, genome-wide association studies (GWAS) have published hundreds of common variants whose allele frequencies are statistically correlated with various illnesses and traits. However, the vast majority of such variants have no established biological relevance to disease or clinical utility for prognosis or treatment."

"More generally, it is now clear that common risk variants fail to explain the vast majority of genetic heritability for any human disease, either individually or collectively (Manolio et al., 2009)."

Until present, nobody has attempted to encode SVs in genomes larger than one base, here we call DIPs. As an example, deep learning has been deployed to predict gene expression from histone modifications [2]. Genome-wide assays of breast cancer by denoising autoencoders (DAs) employs a data-defined learning objective independent of known biology [3].So by using this independent system the information is not captured for any advantage. Convolution neural network has been used for classifying various kinds of tumors [4]. Deep learning has been used for pathology image classification [5] and does not tap into the SV of the genome information for the purpose. Recurrent neural network without deploying SVs of the genome has been used for heart failure onset [6].Articles [7,8] act as a review article for deep learning in pharmaceutical research and drug discovery-SVs of genome for any advantageous role are not mentioned. Brain disorders, such as Alzheimer's disease, are evaluated using brain images using artificial intelligence techniques in article [9], yet heart related disorders use deep learning for magnetic resonance information in [10]. Article [11] tries to make use of transcriptomics data along with deep learning for drug prediction and repositioning, again SVs of the genomic data are not mentioned. Recently in 2019, paper [25], visits the idea of machine learning by SNP-only based approach, which fails to point out the impact of DIPs and its appropriate encoding to facilitate machine and deep learning.

SVs[20] in genomic data are obtained after comparing the patient's DNA sequence with a reference sequence and finding matches and mismatches using tools, such as GenomeBreak [16, 17]. Incorporation of DIPs or InDels to MLCSB cannot be avoided, as we are generating more and more sequences and the data is routinely being downstream analyzed for SVs. As DIPs essentially have all information for CNVs, inversions, translocations and other SVs on genome, encoding them would also indirectly encode the other SVs. Article [18], discusses utilizing tools for these SV detections, then comparing these variations to databases and conducting a knowledge mining [19] where these variations are known to be as-sociated with a disease. DNA sequencing for individuals is becoming increasingly cheaper to obtain, for example via NGS at sequencing centers where it can be done at a scale thereby distributing the fixed cost [15].

## Method

Paper [21], showed qualitatively that the deviation of the sum of the nucleotides in DIPs was generally higher than the deviation of the sum of the nucleotides of the SNPs for the whole genome. In other words, deviations in DIPs were more representative of the individual differences among them and could thus attribute to their differences in endo-phenotype or end-phenotype. As an example, paper [21] took the gender as the end phenotype and showed that the variance (and the standard deviation) between the set of structural variations (DIPs) was much higher than that of the sum of nucleotides of SNPs (Figure 1.), and stating that structural variations were a stronger means to determine the phenotype, i.e. gender here. Inspired by the article, this paper is about fine tuning and quantifying individual DIPs, so we introduce a deviation from consensus score to quantify the differences in SVs for these letters while also using one-hot encoding for the single nucleotide bases.

We developed DMWAS suite to simulate genomic data for genomic co-ordinates as a combination of A, T,G C or a 4-letter combination for a larger letter. Comprised of a Python script genSampleData.py, it can be used to generate genomic variation data specifying quantity of genomic loci, number of patients, frequency of occurrence of DIPs, and the maximum size of a DIP. Details of usage of script are specified in the downloadable ReadMe.md file from GitHub. For illustration purpose we are using 400 genomic coordinates.

In Figure 2 are the simulated data for 40 columns and 8 patients. The simulated genotype data for 200 loci is provided as file multiColumnSample.csv . Once the simulated data is generated, and then we use the script splitMultiColDIPs.py to split each feature column into two columns, one column for 1 letter variants and another column for DIPs variants. The split file is available by name multiColumnSplitSample.csv . In Figure 3 is an example of data with each column doubled as per described method.

With information for the DIPs as second column, we can extract them separately and conduct a clustering of the data getting the divergence score for each DIPs from the consensus. This method of encoding the letters based on divergence from a mean, median or consensus score is called 'DivScoreEncoding'. DivScoreEncoding is different than one-hot, word embedding, index based encoding and other kind of label encoding methods as described in the article 'Text Encoding: A Review' [26].We realize that the InDels or DIPs can be different from each other and the difference in biological relevance such as by means of frameshift of codon or mutation at a point need to be given a score in biological context. The traditional means of encoding texts, such as those discussed at [26], do not take biological evolutionary distance into account when encoding for DIPs or InDels. This method of DivScoreEncoding is applicable to larger insertions and deletions as well as for other SVs in the genome like CNVs, translocations, etc. Cross-species multiple sequence alignment has been tried using phylogenetic tree construction in article [30]. Papers [28,29] generate pathogenic scores of InDels throughout the non-coding genome to classify them into pathogenic or not, and would be clearly very different in terms of method and application, although the similarity remains in terms of the concept of giving a score to the InDels based on a biological role. Figure 4 shows a sample clustering by multiple alignments done words of varying length with consensus and divergence score for each sequence.

For implementing DivScoreEncoding method by clustering, we have chosen T_coffee[1] application to get the divergence score. This third-party software is available online. A wrapper Python script multiColDIPsDiv.py is provided, which automates extraction of the DIPs from multiCoumnSplitSample.csv file, then passes it to T_coffee software for multiple sequence alignment and divergence score determination. Script reverseReadMulti.py is provided to reverse the scores obtained, and script ReplaceMultiColDIPsNew.py can be used to replace the DIPs with appropriate scores. This would lead to file with content such as in figure 5. The resulting file is also provided as MultiColDIPsScored.txt . The Python script enco-deSNPs.py has been provided for this purpose; the resulting final scored and encoded file MultiColDIPsScoredEncoded.txt is also provided. Figure 6 shows a sample scored and encoded file snippet.

Since we had 40 individuals or rows, we generated 40 y-values 0-39, with the 1st row left as that of feature column variable names. File is named as Phenotype.txt . We decided to use several machine learning methods, such as logistic regression, naïve bayes, gradient boosting, bagging, and adaboost, and deploy enhanced form of exhaustive multi-layer perceptron (MLP= in the form of DNN by incorporating early stopping criteria to avoid overfitting, using rectified linear unit (*ReLU*) as activation function to reduce weight adjustment time and addressing the vanishing gradient problem. We also introduce an exhaustive nature of exploration for the right hidden layer and hidden units by varying the number of layers and number of hidden units in the DNN in a loop. Each time the best scores were chosen for its number of hidden layers and units. This exhaustive nature of DNN, when the range was given in realistic bound proved more useful than simply adding hidden layers as in a typical DNN, and thereby gave profound results, so this approach is called 'ExhaustiveDNN'. The scripts ExhaustiveDNN.ipynb and ExhaustiveDNN.py are provided in DMWAS and feeds in Multi-ColDIPsScoredEncoded.txt as input file. The script internally looks for all columns with any null values that are removed before modeling. The data file was also separately checked for Null values and minor allele frequency (MAF) of at least 5% and the resulting encoded file is available at DMWAS as NullMafMultiColDIPsScoredEncoded.txt, which can be used as an alternative. From this file applying F-Test criteria for each of the feature columns we chose the top 1% of the feature set as the final data that the deep and machine learning scripts would work on. Feature set optimization has been an active area of research recently such as what we see in article [31]. Article [27] talks about various applications of deep learning in different spheres of biology, and to which ExhaustiveDNN as part of DMWAS with the DivScoreEncoding methodology can play vital role as it is exhibited in the results section later.

**Results**

ExhaustiveDNN proved very useful. 30% of data was used for test and prediction purpose, results shown as a confusion matrix. In less than a minute it resulted in model that was 100% accurate on the test data with the following configuration of hidden layers and hidden units, and the score on average for each training batch as 96%:hidden units: 8, hidden layers: 2, avg_score:0.9600000023841858. The confusion matrix is shown in figure 7.

Here accuracy is defined as:

**Accuracy = (TP+TN)/(TP+FP+FN+TN)**

Machine learning techniques, mentioned in previous section, were applied as well for which figure 8 shows their corresponding confusion matrices. Each script took less than 1 minute to produce the confusion matrix, precision-recall curve, ROC-curve and list the dominant features. The scripts are available in DMWAS as createLogitReg.py, createAdaBoost.py, createBagging.py, createGradientBoosting.py and createNaiveBayes.py, extratreeclassifier.ipynb, randomforest.ipynb, support vector.ipynb. When there would be imbalance in distribution of cases and control, then PR-Curve metrics would be worthwhile to discuss, as we later plan to scale up the work for larger dataset analysis in future. All results for ROC-curve, PR-Curve, list of dominant column variables, etc. are made available at DMWAS GitHub. Table 1 below summarizes the accuracy values obtained from these machine and deep learning software tools.

Using these approaches, the Naïve Bayes method seems to have the highest positive hits detected with 75% accuracy in this simulated data.

However, using the ExhaustiveDNN approach with a variation of number of layers and hidden units, with early stopping conditions, gave the best result with an accuracy of 100% almost immediately. The trick is to set the initial set of hidden layers and hidden units large enough while running ExhaustiveDNN. The initial opinion of 100% accuracy would be that the model has perhaps done over-fitting, the early stopping condition ensures that over-fitting does not take place. This is further substantiated by the fact that Exhausti-veDNN does not give 100% accuracy in real GTEx data, as discussed later. ExhaustiveDNN when allowed to continue after the 1[st] model has been generated can lead to multiple models, each with different average accuracy score such as that shown below in table 2 at epoch (cycles) of 100. The model with best average score is saved for its configuration to be used on test and real data.

### Application of DMWAS to GTEx V7 Pilot dataset

We used the scripts of DMWAS for Genotype-Tissue Expression (GTEx) project V7 pilot dataset of 185 individuals, for the phenotype coded as MHHRTATT for the people who died of 'heart attack, acute myocardial infarction, acute coronary syndrome', and were able to see that most of the machine learning based algorithms could perform remarkably better for real case data. As an example, Fig. 9 shows the AUC for ROC-curve for logistic regression for the MHHRTATT phenotype. A score of 97.3% accuracy was obtained using logistic regression model of DMWAS as shown through the confusion matrix in Fig 10 for which the test data was taken as the entire GTEx V7 Pilot dataset. The results obtained have been summarized in Table 3. The plots for various confusion matrices are shown as well in Figure 11.

### Discussion

Evidently for the given dataset of GTEx V7 Pilot the methods of support vector and logistic regression substantially outperformed other methods in terms of accuracy. Since true positives in the dataset were significantly less, once scaled up from pilot dataset to whole dataset analysis, such as for GTEx V8 data, the method would help determine the precision rather than just depend on accuracy.

For illustration purpose on real data, we showed how the results improve drastically as we achieve accuracy of 97.3 % for real case data of GTEx V7 Pilot, for logistic regression, compared to only 58.3% as in randomly generated simulated dataset. This 97.3% of accuracy was generated when the entire GTEx V7 Pilot data was used for testing purpose.

The tools and techniques discussed in DMWAS can be applied for solving other data science problems as once the encoding work is completed, the user can use any algorithm of his choice and not be locked into using those provided or suggested in this paper. This applies to the clustering DivScoreEncoding method as well. For instance, we can give each letter a value, then calculate a mean or median score for the complete word, or use other sophisticated clustering method which use fuzzy logic for instance.

**Optimized Feature set for MHHRTATT biomarkers**

These models can help us 'optimize the feature set' i.e., identify dominant variants that are strongly associated to the model - and thus to the disease. The possibility was explored on the simulated dataset as well as prediction was made for MHHRTATT trait for the GTEx V7 pilot dataset to see if we get a score for the DIPs variant columns. Table 4 lists a partial dependence score generated for the simulated data in which the variant columns were also captured. The partial dependence score is calculated for genomic variant columns having single nucleotide variant eg. 214_C would mean the C nucleotide at 214th column in the genome variant file, for showing that just the presence of DIP at a position eg. 232_I means that the 232$^{nd}$ genomic variant column having an insertion, for showing the effect of the insertion variants at that column simply the column number is stated eg. 377. For the real case data the top 10 optimized features were all belonging to InDel class as shown in table 5 for the logistic regression; the column variable name and numbering is as per the GTEx data with extension filename .PED and the actual co-ordinates can be found by looking at the corresponding rows of .MAP file. Note that since the .PED file comprise of one major allele and another minor allele information, the number of columns with regard to the genotype information is twice that of the number of rows in .MAP file and so tracing back of the corresponding genomic map coordinate should be done accordingly. As an example if the optimized feature has name 16,830,168_G, then the .PED file corresponding feature co-ordinate removing the initial 6 columns is 16,830,168 and the genomic variant that is having an effect is G. The corresponding genomic map coordinate line number is CEILING ( 16,830,168 / 4 ) = 4,207,542 . This in the .MAP file corresponds to variant Id kgp30994055 and at position 52587347 of chromosome 23. The list of top associated and least associated genomic variants with their chromosome number, variant Id, and genomic loci are stated in table 5 and table 6 respectively. Apparently, the lowest scoring features were all SNPs (table 6) however the relative difference in the tops scorer and bottom scorer were not huge indicating a rheostat model of combined effect of the variants on the phenotype.

### Conclusion & Future work

This paper has demonstrated and advocates use of clustering divergence score as a new way of genomic variant encoding particularly for structural variants larger than point mutations and demonstrated in for InDels, though the technique can well be applied to other SVs such as copy number variants, etc. Several machine learning algorithms were experimented and MLP (multi-layer perceptron) script with alterations to gain properties of deep learning was developed in Python, such as early stopping condition to avoid overfitting. This led us to 100% accuracy using ExhaustiveDNN for the simulated data while accuracy for the real data was lower; confirming no case of over-fitting as far as the script logic is concerned. Other machine learning techniques such as bagging gave results lower than DNN with highest being 75% using Naïve Bayes for the simulated data. The concept of clustering score is central to the ideas discussed in this paper and once the divergence scores are obtained, the downstream modeling advance algorithm need not be just restricted to those mentioned in DMWAS GitHub page but could also use many other deep and machine learning algorithms such as even genetic algorithm as has been used for GARBO [31].

We gave results for optimized feature for the top 10 genomic variants for MHHRTATT heart disease related death. Future work requires up-scaling the analysis for the entire dataset such as GTEx V8 since the number of cases of individuals having the trait in pilot sample is very limited, thereby having considerable under performance for most of the deep and machine learning models. Future work also asks for exploring and comparing performance of other similar tools that deploy machine and deep learning for GWAS, such as CADD [30] even though it was used in cross-species context, or GARBO[31] which uses fuzzy logic and genetic algorithm, and see if there are complementary aspects that DMWAS can benefit from, in a future version of the tool. The purpose of the current work was to not just describe a method, but also list top genomic variants as-sociated to MHHRTATT. The idea is also to make DMWAS available and deployable for the purpose of deep learning and machine learning application to GWAS.

### Open-Source Development & Supplementary:

The PR-Curve, ROC-Curve, PDValues (partial dependency scores based on the model for the genomic variant columns) for the simulated data and python scripts including script to simulate data is publicly accessible here: *https://github.com/abinarain/DMWAS*. Supplementary materials can be downloaded from https://sites.google.com/a/iitdalumni.com/abi/educational-papers .

### Acknowledgement

The author thanks all developers and teachers of Python programming language. The data was used as part of an authorized access request while the author was employed at A.I. Virtanen Institute employed and working for Genotype-Tissue Expression (GTEx) dataset https://gtexportal.org/home/index.html. The author is grateful to Dr Minna Kaikkonen for giving him the opportunity to work on the data.